%% file: nova_nsi.tex
\RequirePackage{lineno}
\documentclass[aps,prdrc,twocolumn,superscriptaddress,groupedaddress,floatfix,preprintnumbers]{revtex4}

\usepackage{graphicx}  
\usepackage{dcolumn}   
\usepackage{multirow}  
\usepackage{comment}
\usepackage{bm}        
\usepackage{amssymb}   
\usepackage{amsmath}
\usepackage{booktabs} 
\usepackage{braket}
\usepackage{breqn}

\usepackage{float}      
\usepackage{caption}
\usepackage{subcaption}

\usepackage[normalem]{ulem}
\usepackage{siunitx}

\usepackage{xcolor}
\usepackage{xspace, tabularx}
\usepackage{caption}
\usepackage{threeparttable}

\newcommand{\deltacp}
{\ensuremath{\delta_{\tiny\text{CP}}}\xspace}
\newcommand{\epsab}{\ensuremath{\varepsilon_{\alpha\beta}}\xspace}
\newcommand{\deltaab}{\ensuremath{\delta_{\alpha\beta}}\xspace}
\newcommand{\epsetau}{\ensuremath{\varepsilon_{e\tau}}\xspace}
\newcommand{\deltaetau}{\ensuremath{\delta_{e\tau}}\xspace}
\newcommand{\epsemu}{\ensuremath{\varepsilon_{e\mu}}\xspace}

\newcommand{\epsmutau}{\ensuremath{\varepsilon_{\mu\tau}}\xspace}

\newcommand{\beq}{\begin{equation}}
\newcommand{\eeq}{\end{equation}}
\newcommand{\beqa}{\begin{eqnarray}}
\newcommand{\eeqa}{\end{eqnarray}}

\newcommand{\nue}{\ensuremath{\nu_{e}}\xspace}

\newcommand{\numu}{\ensuremath{\nu_{\mu}}\xspace}
\newcommand{\nutau}{\ensuremath{\nu_{\tau}}\xspace}

\begin{document}

\preprint{FERMILAB-PUB-24-0108-PPD}
\title{Search for $CP$-Violating Neutrino Nonstandard Interactions with the NOvA Experiment}

\input novansi2023.tex


\begin{abstract}
This Letter reports a search for charge-parity ($CP$) symmetry violating nonstandard interactions (NSI) of neutrinos with matter 
using the NOvA Experiment, and examines their effects on the determination of the standard oscillation parameters. Data from $\nu_{\mu}(\bar{\nu}_{\mu})\rightarrow\nu_{\mu}(\bar{\nu}_{\mu})$ and $\nu_{\mu}(\bar{\nu}_{\mu})\rightarrow\nu_{e}(\bar{\nu}_{e})$ oscillation channels are used to measure the effect of 
the NSI parameters $\varepsilon_{e\mu}$ and $\varepsilon_{e\tau}$. With 90\% CL the magnitudes of the NSI couplings are constrained to be $|\epsemu| \, \lesssim 0.3$ and $|\epsetau| \, \lesssim 0.4$. A degeneracy at $|\varepsilon_{e\tau}| \, \approx 1.8$ is reported, and we observe that the presence of NSI limits sensitivity to the standard $CP$ phase \deltacp.
\end{abstract}

\maketitle

Theoretical and experimental research extending over many decades has yielded an established framework for neutrino oscillation phenomena \cite{Pontecorvo:1957cp,ref:Pontecorvo1958,Maki:1962mu,Pontecorvo:1967fh,Fritzsch:1975rz,Bilenky:1978nj,Super-Kamiokande:1998kpq,Super-Kamiokande:2002ujc,SNO:2002hgz,KamLAND:2002uet,MINOS:2006foh,T2K:2011ypd,DoubleChooz:2011ymz,DayaBay:2012fng,RENO:2012mkc}. 
According to this framework, neutrino flavor eigenstates ($\nu_e, \nu_{\mu}, \nu_{\tau}$) are a mixture of three mass eigenstates ($\nu_1, \nu_{2}, \nu_{3}$), such that 
\begin{linenomath*}
\begin{equation}\label{eq_nu_mixing}
    \nu_{\alpha} = \sum_i U_{\alpha i}\nu_i; \quad \alpha = (e,\mu,\tau); \quad i = (1,2,3),
\end{equation}
\end{linenomath*}
where $U$ is the Pontecorvo-Maki-Nakagawa-Sakata $3\times 3$ mixing matrix \cite{Maki:1962mu,Pontecorvo:1967fh}. $U$ is parametrized in terms of three mixing angles, $(\theta_{12}$, $\theta_{13}$, $\theta_{23})$, and a charge-parity ($CP$) symmetry violating phase, \deltacp, the measurement of which is of paramount relevance.
If \deltacp is not an integer multiple of $\pi$, it would indicate $CP$ violation in the neutrino sector providing a feasible explanation to open questions in cosmology, including the observed matter-antimatter asymmetry in the Universe \cite{Pilaftsis:1997jf,Buchmuller:2004nz,Buchmuller:2005eh}. 
Accurate values of the mixing angles $\theta_{12}$ and $\theta_{13}$ have been obtained by solar \cite{Cleveland:1998nv,SNO:2002hgz,SNO:2003bmh,SNO:2005oxr,Super-Kamiokande:2002ujc,Super-Kamiokande:2003yed} and reactor \cite{KamLAND:2002uet,DoubleChooz:2011ymz,DayaBay:2012fng,RENO:2012mkc} experiments. A precise measurement of the mixing angle $\theta_{23}$, which determines the coupling of the $\numu$ and $\nutau$ states to $\nu_3$, is part of current and future research efforts \cite{NOvA:2021nfi,T2K:2023smv,DUNE:2020jqi,Hyper-Kamiokande:2022smq}.

The frequency of neutrino oscillations is mainly governed by the mass-squared splittings $\Delta m^2_{ji} \equiv m_{j}^2 - m_{i}^2$ between two neutrino mass eigenstates $\nu_{j,i}$. Measurements from solar neutrino experiments \cite{Super-Kamiokande:2002ujc,SNO:2002hgz,SNO:2003bmh,KamLAND:2008dgz,Super-Kamiokande:2008ecj,Super-Kamiokande:2010tar,SNO:2011hxd,Super-Kamiokande:2016yck} determined that $\Delta m^2_{21}$ is positive, while the sign of $\Delta m^2_{32}$ is still unknown. Taking the mass state $\nu_1$ as having the largest contribution from the flavor state $\nu_e$, if $\Delta m^2_{32} > 0$, neutrinos are said to have a normal mass ordering (NO), while if $\Delta m^2_{32} < 0$, neutrinos would have an inverted mass ordering (IO). 

Standard interactions of neutrinos with matter change the oscillation probability as compared to in vacuum. This Mikheyev-Smirnov-Wolfenstein effect \cite{Wolfenstein:1977ue,Mikheyev:1985zog,Mikheev:1986wj} is produced by the coherent forward scattering of neutrinos on electrons in the Earth's crust and enhances (suppresses) the rate of $\numu \to \nue$ ($\bar{\nu}_{\mu} \to \bar{\nu}_e$) for the NO, with a reversed effect for the IO.

The current uncertainties of some of the oscillation parameters keep alive the possibility of new-physics scenarios predicting neutrino oscillations together with additional subleading phenomena, including, but not limited to, increased number of flavor neutrinos, neutrino decay, and quantum decoherence. A precise measurement of the oscillation parameters is crucial to probe these alternative models, improving our understanding of neutrino physics, consequently. Of particular interest to this Letter, the extension of the above-described framework proposes the presence of subleading neutrino interactions with matter, and has been of interest to the community because of the large potential impact the new interactions may have on the determination of the oscillation parameters \cite{Kopp_2008,Gonzalez-Garcia:2011vlg,Friedland:2012tq,deGouvea:2016pom,Agarwalla:2016fkh}. These kinds of models also have been considered to explain recent differences between individual measurements of \deltacp between NOvA and T2K \cite{Denton:2020uda,Chatterjee:2020kkm}. Given NOvA's enhanced matter effect, it is expected for these phenomena to be dominant on NOvA data, greatly contributing to the conclusions presented in \cite{Denton:2020uda,Chatterjee:2020kkm}. This Letter considers such a possibility in which, in particular, neutral currentlike nonstandard interactions (NSIs) of neutrinos with matter are included (charged currentlike NSIs, affecting the production and detection of neutrinos, are not considered here). These neutral currentlike NSIs can be expressed by an effective four-fermion Lagrangian,
\begin{linenomath*}
\begin{equation}\label{eq_Lagrangian}
    \mathcal{L} = -2\sqrt{2} G_F \varepsilon^{fX}_{\alpha\beta} 
    \left( \bar{\nu}_{\alpha}\gamma^{\mu}P_L\nu_{\beta} \right)
    \left( \bar{f}\gamma_{\mu}P_Xf \right),
\end{equation}
\end{linenomath*}
where $G_F$ is the Fermi constant, $P_X$ is the left ($X=L$) or right ($X=R$) chirality projection operator, and $\varepsilon^{fX}_{\alpha\beta}$ are dimensionless coefficients quantifying the strength of the NSIs between neutrinos of flavor $\alpha$ and $\beta$ and the matter field $f$, relative to the weak scale~\cite{Farzan:2017xzy}. 
If $\varepsilon^{fX}_{\alpha\beta} \neq 0$, beyond standard model physics would manifest in the form of lepton flavor violation ($\alpha \neq \beta$) and/or lepton flavor universality violation ($\alpha = \beta$).
For neutrinos traveling though the Earth, the interaction with matter can be parametrized in terms of the effective NSI couplings $\epsab$, which encompass the new beyond standard model phenomenology \cite{Farzan:2017xzy}.

In this scenario, where the phenomenology of neutrino propagation in matter is altered by the presence of NSIs, the Hamiltonian is modified to include the effective parameters governing the new interactions in the standard matter potential matrix,  
leading to 

\begin{linenomath*}
\begin{equation}\label{eq_Hamiltonian}
H =
 U \left( 
        \begin{array}{ccc}
            0 & 0 & 0 \\
            0 & \Delta_{21} & 0 \\
            0 & 0 & \Delta_{31}
        \end{array}    
    \right)U^{\dagger} + V \left(
        \begin{array}{ccc}
            1 + \varepsilon_{ee} & \varepsilon_{e\mu} & \varepsilon_{e\tau} \\
            \varepsilon_{e\mu}^* & \varepsilon_{\mu\mu} & \varepsilon_{\mu\tau} \\
            \varepsilon_{e\tau}^* & \varepsilon_{\mu\tau}^* & \varepsilon_{\tau\tau} \\
        \end{array}
    \right),
\end{equation}
\end{linenomath*}
where $\Delta_{ji} \equiv \Delta m^2_{ji}/2E$, $E$ is the neutrino energy, and $V = \sqrt{2}G_FN_e$ corresponds to the normal matter potential leading to the Mikheyev-Smirnov-Wolfenstein effect, with $N_e$ and $G_F$ being the electron number density and the Fermi coupling constant, respectively \cite{Miranda:2015dra}. In Eq.~(\ref{eq_Hamiltonian}) we have written the effective NSI parameters by summing over the matter fields $f$, considering their contribution weighted to the effective density relative to the electron one. The parameter $V$ can be written as a function of the matter density $\rho$,
\begin{linenomath*}
\begin{equation}\label{eq:MattPotential}
    V \simeq 7.6 ~Y_e \times 10^{-14} \left(\frac{\rho}{\rm{g/cm}^3}\right) \, \rm{eV},
\end{equation}
\end{linenomath*}
with $Y_e = N_e/(N_p + N_n) \simeq 0.5$, the relative electron number density in the Earth's crust.

\begin{figure}[ht]
    \includegraphics[width=\linewidth]{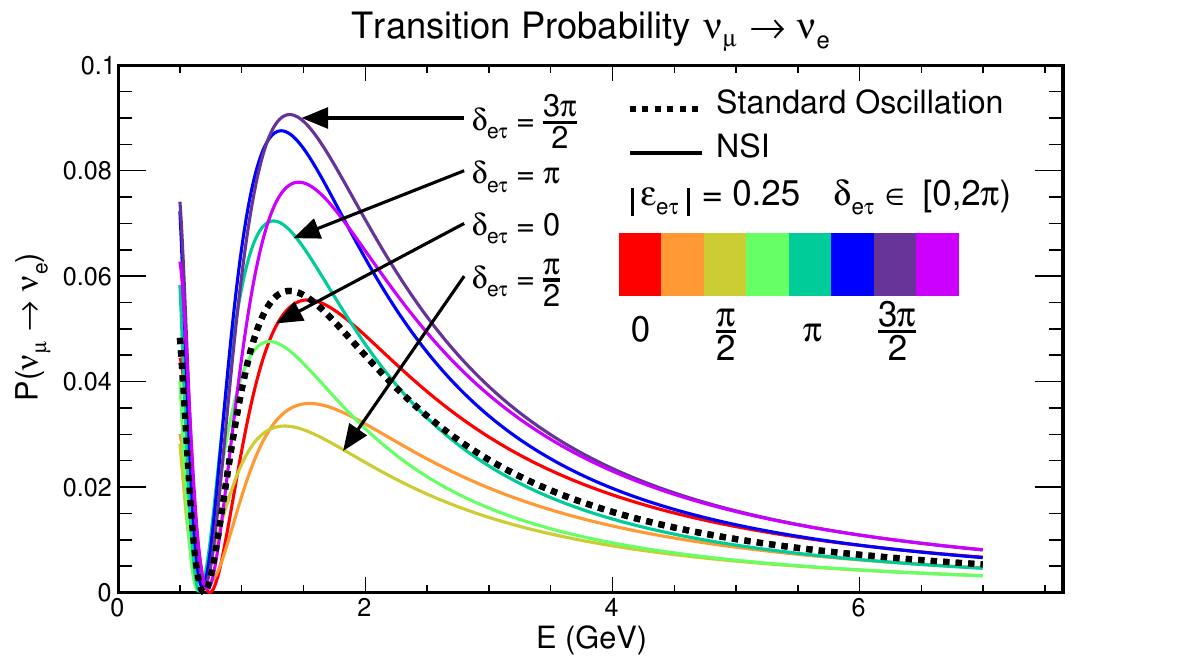}
    \caption{
Oscillation probability $P(\nu_{\mu}\to \nu_{e})$ for neutrinos traveling 810 km through the Earth. 
The standard oscillation prediction (black dashed line) is compared with the NSI model with $\left|\epsetau\right|=0.25$ and different values of its corresponding phase, $\delta_{e\tau}$. The standard three-flavor oscillation parameters are set to the best-fit values reported in Ref.~\cite{NOvA:2021nfi}: $\Delta m^2_{32} = +2.41\times 10^{-3}$ eV$^2$, $\sin^2\theta_{23} = 0.57$, $\deltacp = 0.82\,\pi$.
    }
    \label{fig:oscProb}
\end{figure}

The NSI couplings are, in general, complex quantities and can be written as
\begin{linenomath*}
\begin{equation}\label{eq_nsi_param}
    \varepsilon_{\alpha\beta} = \left|\varepsilon_{\alpha\beta}\right|e^{i\delta_{\alpha\beta}},
\end{equation}
\end{linenomath*}
where $\delta_{\alpha\beta}$ are new $CP$-violating phases associated with each NSI amplitude. However, due to the Hermiticity of the Hamiltonian in Eq.~(\ref{eq_Hamiltonian}), $\varepsilon_{\alpha\beta}=\varepsilon_{\beta\alpha}^{*}$, implying that the on-diagonal terms are real. 
For the off-diagonal couplings, the phases $\delta_{\alpha\beta}$ along with the moduli $\left|\varepsilon_{\alpha\beta}\right|$ must be considered. As an example, the impact of the NSI phase $\delta_{e\tau}$ on  
the \nue flavor appearance at the NOvA baseline can be seen in Fig.~\ref{fig:oscProb}.

The effect of NSIs on neutrino phenomenology has been studied by several experiments.
The MINOS Collaboration reported a 90\% CL allowed region on $|\epsetau|$ strongly dependent on the effective $CP$ phase $\deltacp + \delta_{e\tau}$ \cite{MINOS:2016sbv}.

More recently, using high energy atmospheric neutrino data, the IceCube Collaboration has set stringent constraints (90\% CL) of $|\epsemu|\;\le 0.146$, and $|\epsetau|\;\le 0.173$ 
\cite{IceCubeCollaboration:2021euf}, and $-0.0041 < \epsmutau < 0.0031$ when accounting for an effective real-valued parameter \cite{IceCube:2022ubv}. However, unlike MINOS and NOvA, IceCube is insensitive to \deltacp, so their analysis assumes a fixed \deltacp = 0.


NOvA \cite{ref:NOvADesign2007} is a long-baseline experiment designed to measure neutrino oscillations through the disappearance of $\nu_{\mu}$ ($\bar{\nu}_{\mu}$) and the appearance of $\nu_{e}$ ($\bar{\nu}_{e}$) from a beam mainly composed of $\nu_{\mu}$ ($\bar{\nu}_{\mu}$) \cite{NOvA:2021nfi}. Muon (anti)neutrinos are produced in the NuMI beam \cite{ref:NuMI2016} through the decay of pions and kaons resulting from 120 GeV protons scattering off a fixed carbon target. Charged pions and kaons, focused by two magnetic horns, decay into $\mu^{+}(\mu^{-})$ and $\numu(\bar{\nu}_{\mu})$. The polarity of the horns is used to select between a neutrino$\,$(antineutrino) dominated beam composed of $93\%\,\,(92\%)$ pure $\nu_{\mu}\,(\bar{\nu}_{\mu})$  \cite{NOvA:2021nfi}. Neutrinos are detected by two functionally identical tracking calorimeters. The near detector is located at Fermilab, 100~m underground and $\sim$1~km from the neutrino production target, while the far detector is placed $\sim$810~km away from the neutrino source at Ash River, Minnesota.
As the FD is on the surface, it receives a cosmic-ray flux of $\sim$130~kHz. NOvA detectors are placed 14.6~mrad off-axis with respect to the beamline, producing a narrow-band energy spectrum at the FD peaked around 1.8~GeV, which 
optimizes standard three-flavor oscillations for the NOvA baseline. A detailed description of the NOvA detectors can be found elsewhere \cite{ref:NOvADesign2007,NOvA:2018gge}.

The sensitivity of NOvA to \deltacp and the sign of $\Delta m^2_{32}$ is strongly related to standard matter effects on electron (anti)neutrinos as they travel through the Earth. Because of degeneracies, the presence of NSI $\varepsilon_{\alpha\beta}$ terms could complicate the measurement of standard oscillation parameters. Here, we analyze the same dataset presented in Ref.~\cite{NOvA:2021nfi} corresponding to an equivalent exposure of $13.6 \times 10^{20}$~protons on target (POT equiv.) in $\nu$ mode from \SI{555.3}{\second} of integrated beam-pulse time recorded from February 6, 2014, to March 20, 2020, and $12.5 \times 10^{20}$ protons on target (POT) in $\bar{\nu}$ mode delivered during \SI{321.1}{\second} of integrated beam-pulse time recorded from June 29, 2016, to February 26, 2019. 

The specific aspects of the analysis framework are described in more detail in Ref.~\cite{NOvA:2021nfi}, including event simulation, selection, and reconstruction criteria. The analyzed data and the treatment of systematic uncertainties also entirely follow the aforementioned reference. 

Since recent measurements with atmospheric neutrinos reached a stronger sensitivity on \epsmutau \cite{IceCubeCollaboration:2021euf}, we do not include this coupling in the present study.
This analysis focuses on the parameters \epsemu and \epsetau, which modify the FD electron neutrino appearance probability,  where NOvA has competitive sensitivity.

For the analysis presented here, predicted energy spectra are constructed by varying the standard and nonstandard oscillation parameters. These spectra are compared to NOvA FD data using a Poisson negative log-likelihood ratio, $-2\text{ln}\mathcal{L}$. 
Systematic uncertainties are included as nuisance parameters and assigned penalty terms equal to the square of the number of standard deviations by which they vary from their nominal values.
The parameter values that minimize the $-2\text{ln}\mathcal{L}$ are taken as the best fit. During the fit, in addition to all Pontecorvo-Maki-Nakagawa-Sakata oscillation parameters, only one NSI \epsab coupling is taken to be nonzero at a time, with all other sectors neglected. The leading-order dependence is not $\deltaab$ alone, but instead the sum of phases $\deltacp + \deltaab$. As in the MINOS analysis \cite{MINOS:2016sbv}, the measurement of the NSI strength $|\varepsilon_{\alpha\beta}|$ is done with respect to this sum, where the result is profiled over the difference $\deltacp - \deltaab$.
Using the sum takes into account the degeneracy in phases, while profiling over the difference allows us to consider all linearly independent combinations of \deltaab and \deltacp. 
The fitter repeatedly finds the local best fits starting from randomly seeded combinations of all fitted parameters until it determines that the global best fit has very likely been found \cite{Raskutti:2013}.

As NOvA is less sensitive to $\sin^2\theta_{12}$, $\Delta m^2_{21}$, and $\sin^2\theta_{13}$ compared to solar and reactor experiments, in contrast to the standard three-flavor analysis \cite{NOvA:2021nfi}, we use high-precision external measurements from Refs.~\cite{DayaBay:2018yms,RENO:2018dro,DoubleChooz:2019qbj,KamLAND:2013rgu} for these parameters in the form of a Gaussian constraint in the fit. 
The global averages are limited to reactor-only experiments where the baseline is too short for NSIs to play a large role in the result. This avoids possible contamination from the effects of NSIs in results that are measured assuming a standard oscillation model.

An additional consideration is that the strength of the NSI signal is proportional to the matter density, $\rho$. As the neutrinos from the NuMI beam propagate up to a depth of 11~km underground, they experience different rock densities. This is taken into account during the fit by treating $\rho$ as a nuisance parameter. The oscillation model predicts no difference in using the average density compared to using slices of density, hence the average is used.
To estimate the average matter density and its uncertainty, the {\scriptsize CRUST}1.0 model \cite{Laske:2000crust,laske:2013update} is compared to deep bore datasets \cite{Gorbatsevich:2014, Beyer:1978} as well as samples from the NOvA near detector site. A value of $\langle\rho\rangle=\left(2.74 \pm 0.10\right)~\rm{g/cm}^3$ is implemented as a Gaussian constraint in the fit. 

\begin{figure}[t!]
    \makebox[\columnwidth][l]{
        {\label{fig:numu-spectra}\includegraphics[width=0.5\linewidth]{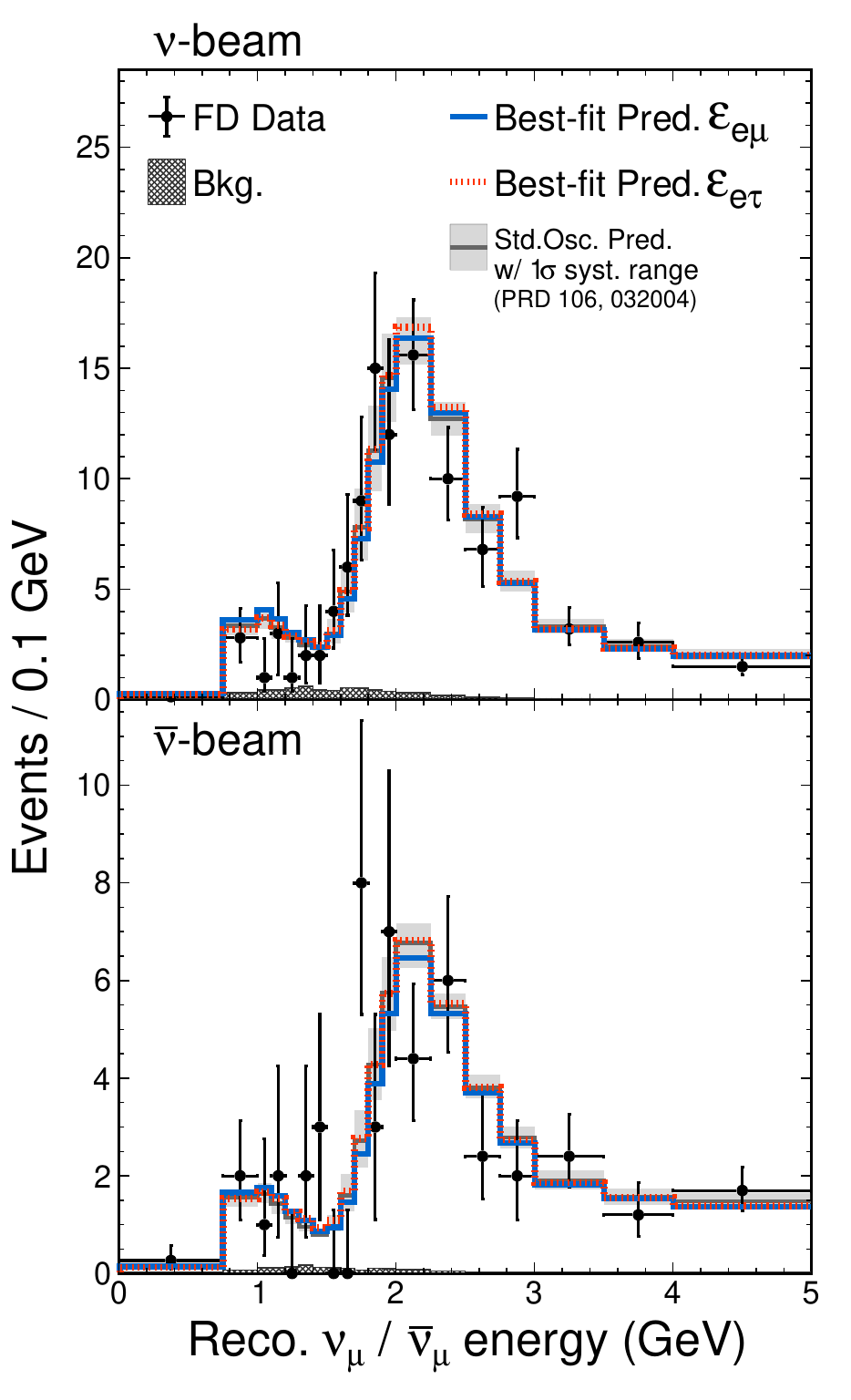}}
        {\label{fig:nue-spectra}\includegraphics[width=0.5\linewidth]{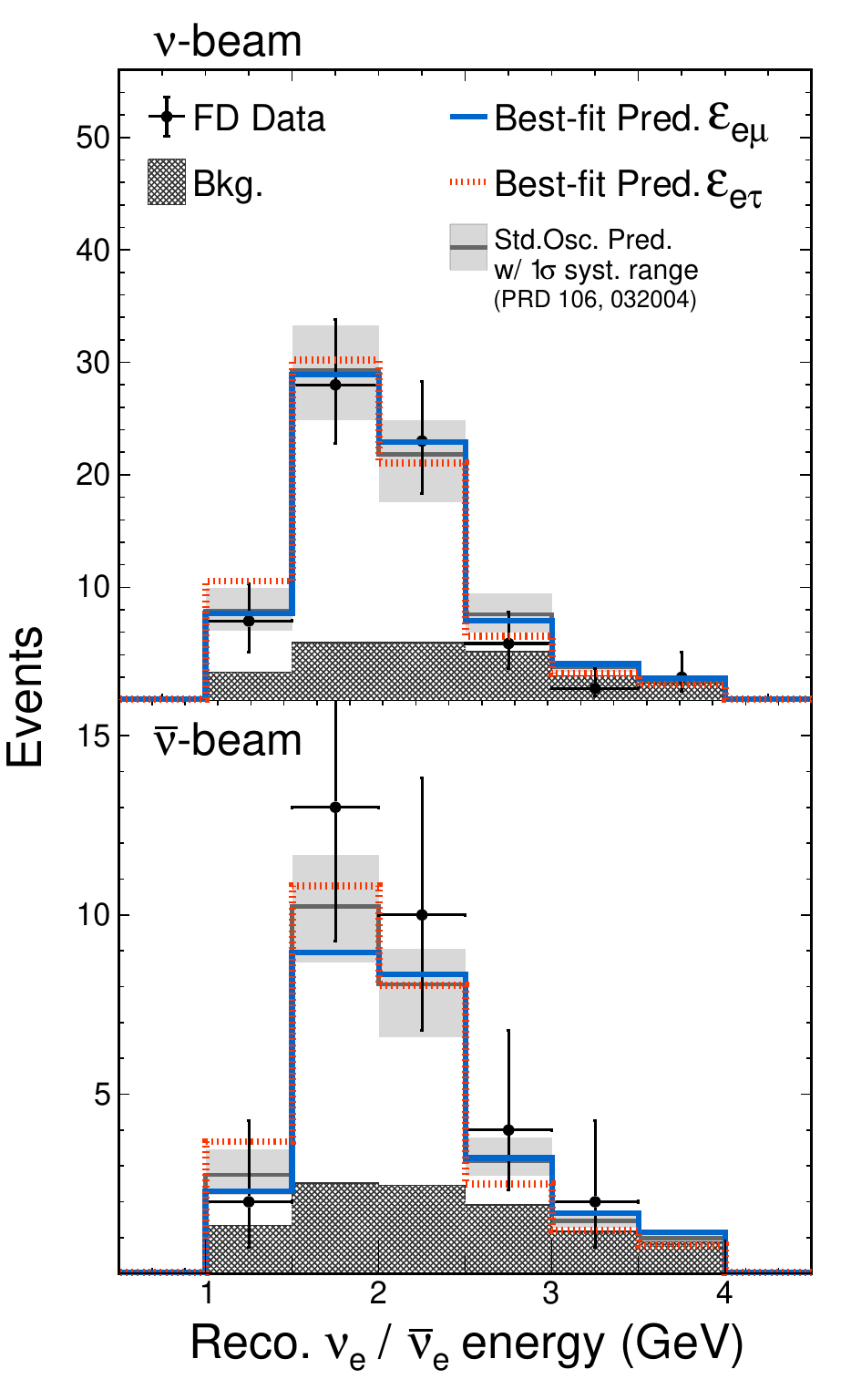}}
    }
    \caption{Reconstructed FD energy spectra for $\nu_{\mu}(\bar{\nu}_{\mu}$)-CC events (left) and $\nu_{e}(\bar{\nu}_{e}$)-CC events (right) from a predominantly neutrino (top) or antineutrino (bottom) beam, with predicted background shown in gray. The standard oscillation prediction (solid black histogram) and its corresponding 1$\sigma$ systematic uncertainty range are compared with the best-fit predictions of this analysis for \epsemu (solid blue line) and \epsetau (dotted red line).}
    \label{fig:fd-spectra}
\end{figure}
\begin{figure}[t!]
        \centering
        \includegraphics[width=\linewidth]{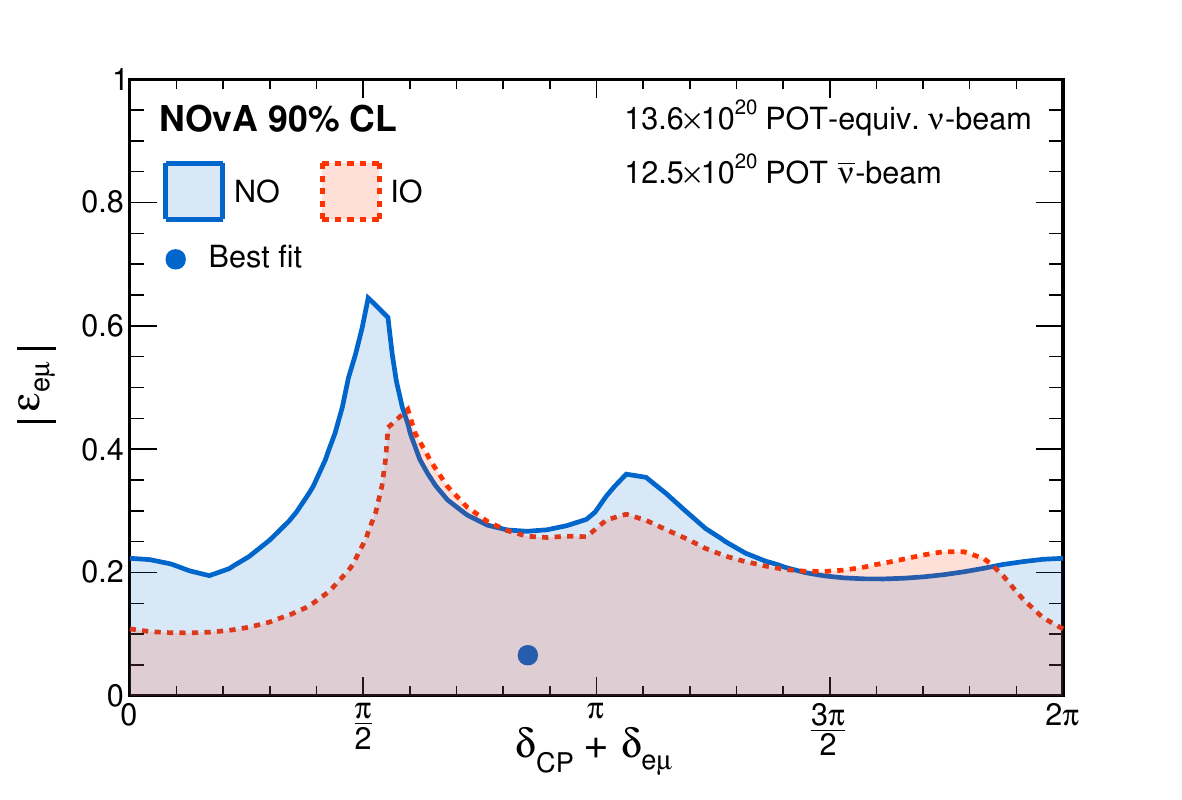}
        \caption{NOvA 90$\%$ CL allowed region for the $\varepsilon_{e\mu}$ vs.~$(\deltacp + \delta_{e\mu})$ parameter space, for NO (solid blue) and IO (dashed red). The global best fit is found at the NO hypothesis.}
        \label{figure:result_emu}
\end{figure}
\begin{figure}[t!]
    \centering
    \includegraphics[width=\linewidth]{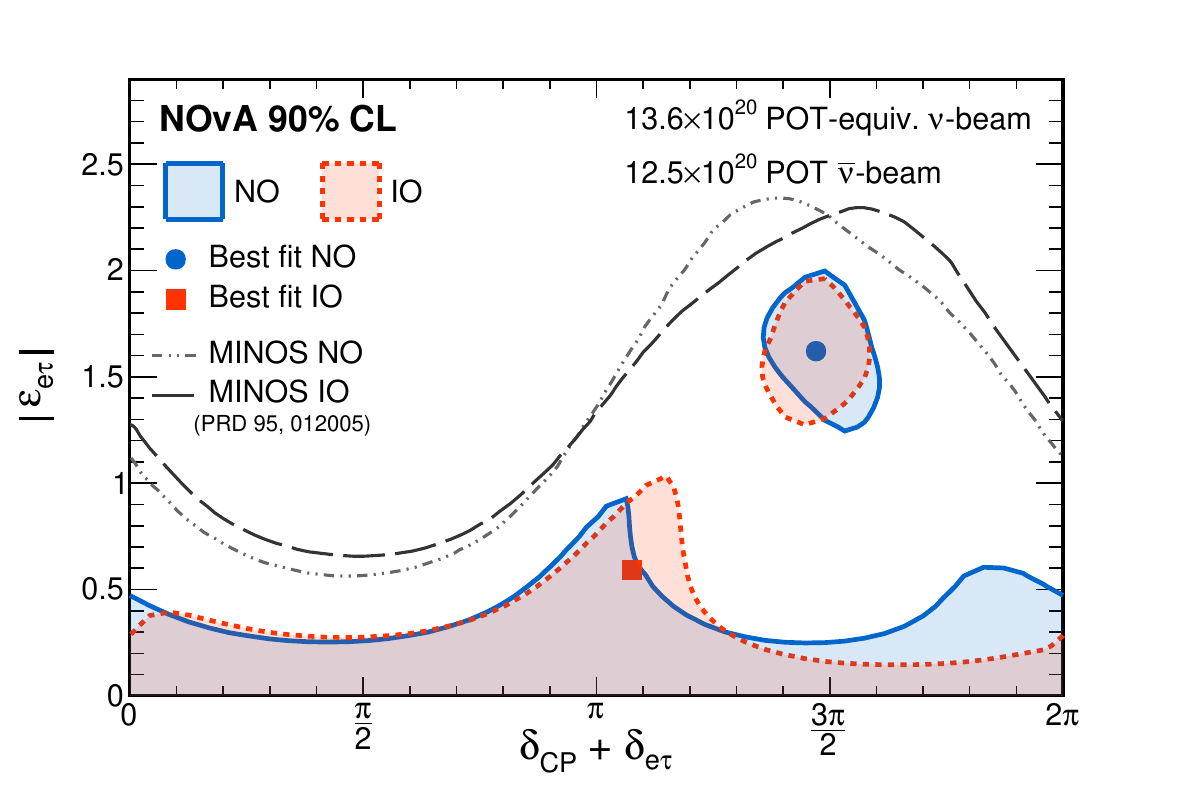}
    \caption{NOvA 90$\%$ CL allowed region for the $\left|\varepsilon_{e\tau}\right|$ vs.~$(\deltacp + \delta_{e\tau})$ parameter space, for NO (solid blue) and IO (dashed red). Both data fits for NO and IO are found degenerate at the same $-2\text{ln}\mathcal{L} = 172.9$. No mass hierarchy preference is observed. MINOS contours \cite{MINOS:2016sbv} are included for comparison.}
    \label{figure:result_etau}
\end{figure}

The observed numbers of $\numu \, (\bar{\nu}_{\mu})$ and $\nue \, (\bar{\nu}_{e})$ events at the NOvA FD are shown in Fig.~\ref{fig:fd-spectra} together with the standard oscillations (SO) best-fit prediction, and its corresponding estimated background \cite{NOvA:2021nfi}. The largest contribution to backgrounds for FD $\nue \, (\bar{\nu}_{e})$ events comes from beam-produced $\nue \, (\bar{\nu}_{e})$ components, on which the addition of the NSI parameters studied here only have a marginal impact. Figure \ref{fig:fd-spectra} also exhibits the best-fit predictions obtained in this analysis (including the SO parameters and the corresponding NSI parameter), showing that NSI predictions are consistent with the spectra produced in the SO model within uncertainty. Figures \ref{figure:result_emu} and \ref{figure:result_etau} constitute the main results of this analysis, showing the 90\% CL allowed regions for the corresponding NSI parameter space. No evidence for NSIs is observed at $90\%\,\text{CL}$, and we place constraints on the absolute value of the NSI couplings \epsemu and \epsetau, for both neutrino mass orderings. These results do not use the unified approach of Feldman and Cousins \cite{ref:FeldmanCousins1998}, as preliminary tests on randomly Feldman and Cousins corrected oscillation parameter bins indicated that Wilks' theorem \cite{Wilks:1938dza} is satisfied, from where frequentist CL intervals are constructed by taking the best-fit $-2\text{ln}\mathcal{L}$ as reference.

The best fit for the \epsemu sector is found at NO with $-2\text{ln}\mathcal{L} = 173.3$ (compared with the SO best fit, $-2\text{ln}\mathcal{L} = 173.55$ \cite{NOvA:2021nfi}) for 170 degrees of freedom, with the IO hypothesis disfavored at $0.6\,\sigma$. For the \epsetau analysis, no significant preference ($\Delta\chi \lesssim 0.034$) is observed for the mass ordering, and the NO and IO hypotheses best fits are found to have $-2\text{ln}\mathcal{L}$ = 172.9. For the \epsetau sector (Fig.~\ref{figure:result_etau}), a degeneracy allowing large values of this NSI amplitude is also observed, similar to what was previously obtained by MINOS \cite{MINOS:2016sbv} (area under the black lines in Fig.~\ref{figure:result_etau}), but here NOvA is excluding a large portion of that reported allowed region. The origin of this upper contour can be explained as follows: changing $\epsetau$ changes the reconstructed spectra in distinct ways for $\nu_{e}$ and $\bar{\nu}_{e}$. Based on the joint analysis reported in Ref.~\cite{NOvA:2021nfi}, the $\nu$--$\bar{\nu}$ fit cancels most of the opposite effects induced by possible NSIs on neutrinos and antineutrinos. However, at $\epsetau = 1.8$ and $\deltacp + \deltaetau = \frac{3\pi}{2}$, both the $\nu_{e}$ and $\bar{\nu}_{e}$ spectra are identical to the SO ($\epsetau = 0$) spectra, creating a degenerate region around that point. As the data are consistent with the SO spectra, this point is also allowed. 
It is worth noticing that some phenomenological treatments elude strong NSI couplings such as \epsetau $> 1.0$ since it would elevate the contribution of a (otherwise negligible) number of terms in the full $\nu_{\mu}\to \nu_{e}$ oscillation amplitude \cite{PhysRevD.86.113015}. Nonetheless, our analysis presents new experimental evidence that improves constraints on previously allowed large \epsetau. Analyzing a wider range of neutrino energies, and possibly combining with measurements from other experiments, is being explored to increase sensitivity to the upper contour in the future.

\begin{figure}[t!]
    \includegraphics[width=\linewidth]{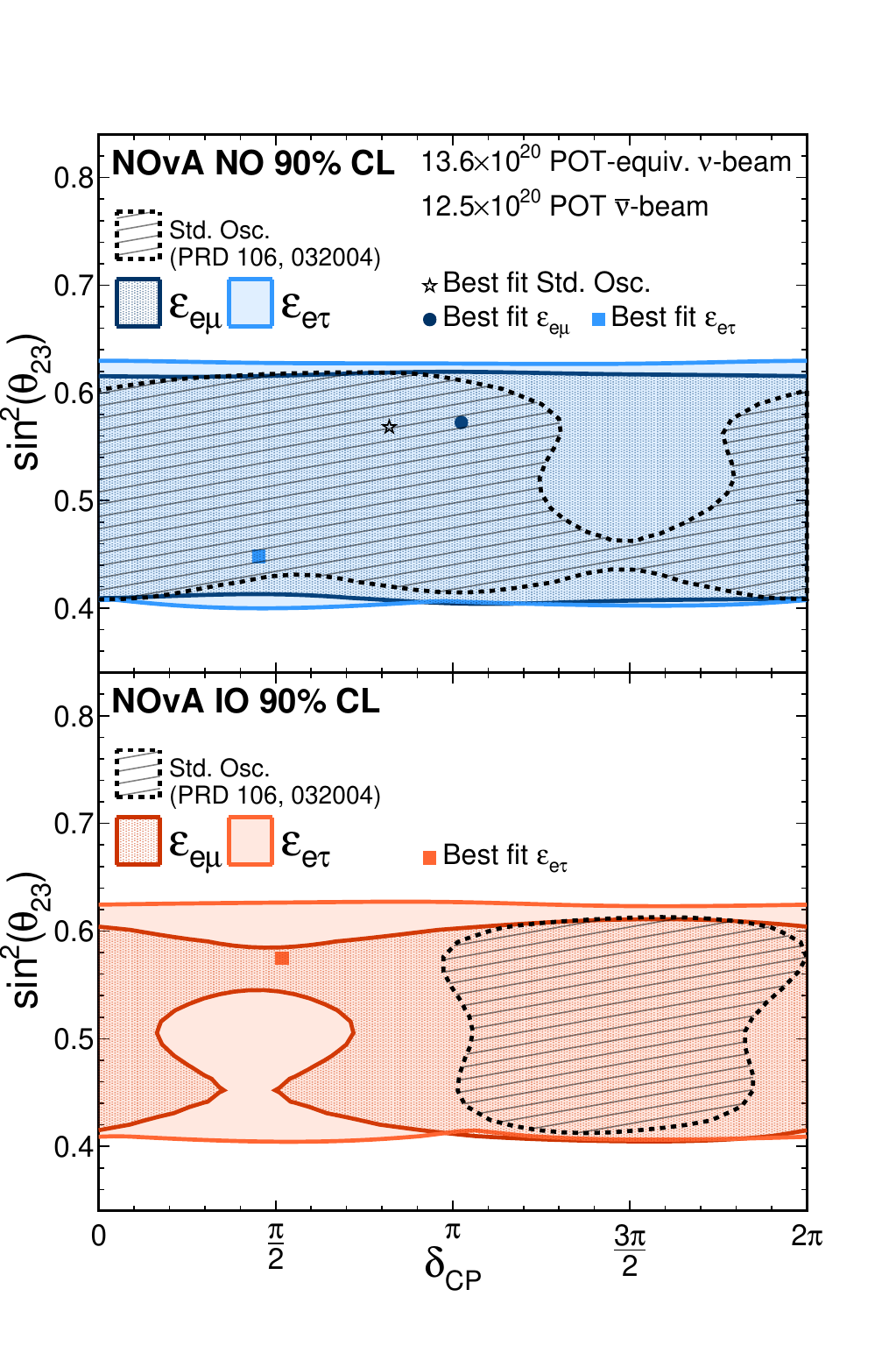}
    \caption{NOvA 90$\%$ CL allowed region for the $\sin^2\theta_{23}$ vs.~\deltacp parameter space, for NO (top) and IO (bottom), comparing the latest NOvA standard oscillation result (dashed shaded black region)~\cite{NOvA:2021nfi} with cases when the NSI sectors $\epsemu$ (solid dark blue and orange) or $\epsetau$ (solid light blue and orange) are included in the fit.}
    \label{fig:ssth23_deltaCP}
\end{figure}

The potential presence of NSI may decrease the sensitivity to the SO $CP$ phase \deltacp. To study this effect, we construct the SO parameter space reported in the previous result from NOvA, allowing for the effects of $|\varepsilon_{\alpha\beta}|$ and its phase $\delta_{\alpha\beta}$. There is a degeneracy between \deltacp and $\delta_{\alpha\beta}$ that this analysis does not break. As Fig.~\ref{fig:ssth23_deltaCP} illustrates, the sensitivity to \deltacp is weakened for both neutrino mass orderings, while the constraints on $\sin^2\theta_{23}$ are scarcely modified.

In summary, we have performed an analysis of the NOvA FD data within a complete three-neutrino framework including NSI couplings that may affect the way neutrinos change their flavor. Including the NSI parameters $\epsemu$ and $\epsetau$ separately only marginally improves the fit to data, indicating the NOvA data alone can be explained by a standard three-neutrino model. The analysis constrains the NSI amplitudes to $|\epsemu| \, \lesssim 0.3$ and $|\epsetau| \, \lesssim 0.4$, although an allowed region for large values of $|\epsetau|$ is observed due to degeneracies between the parameters. The bounds on \epsetau are tighter than previous limits reported by \cite{MINOS:2016sbv}, and both bounds narrow the parameter space available to explain observed discrepancies between neutrino oscillation experiments. We also observe that the possible existence of NSI would affect the measurement of the Dirac $CP$-violating phase, which highlights the need for future experiments to break this degeneracy to further probe the potential for new physics in the neutrino sector.

\textit{Acknowledgments}---This document was prepared by the NOvA Collaboration using the resources of the Fermi National Accelerator Laboratory (Fermilab), a U.S. Department of Energy, Office of Science, HEP User Facility. Fermilab is managed by Fermi Research Alliance, LLC (FRA), acting under Contract No. DE-AC02-07CH11359. This work was supported by the U.S. Department of Energy; the U.S.~National Science Foundation; the Department of Science and Technology, India; the European Research Council; the MSMT CR, GA UK, Czech Republic; the RAS, MSHE, and RFBR, Russia; CNPq and FAPEG, Brazil; UKRI, STFC and the Royal Society, United Kingdom; and the state and University of Minnesota.  We are grateful for the contributions of the staffs of the University of Minnesota at the Ash River Laboratory, and of Fermilab.

\bibliography{nova_nsi}

\end{document}

%% file: novansi2023.tex
\newcommand{\ANL}{Argonne National Laboratory, Argonne, Illinois 60439, 
USA}
\newcommand{\Bandirma}{Bandirma Onyedi Eyl\"ul University, Faculty of 
Engineering and Natural Sciences, Engineering Sciences Department, 
10200, Bandirma, Balıkesir, Turkey}
\newcommand{\ICS}{Institute of Computer Science, The Czech 
Academy of Sciences, 
182 07 Prague, Czech Republic}
\newcommand{\IOP}{Institute of Physics, The Czech 
Academy of Sciences, 
182 21 Prague, Czech Republic}
\newcommand{\Atlantico}{Universidad del Atlantico,
Carrera 30 No.\ 8-49, Puerto Colombia, Atlantico, Colombia}
\newcommand{\BHU}{Department of Physics, Institute of Science, Banaras 
Hindu University, Varanasi, 221 005, India}
\newcommand{\UCLA}{Physics and Astronomy Department, UCLA, Box 951547, Los 
Angeles, California 90095-1547, USA}
\newcommand{\Caltech}{California Institute of 
Technology, Pasadena, California 91125, USA}
\newcommand{\Cochin}{Department of Physics, Cochin University
of Science and Technology, Kochi 682 022, India}
\newcommand{\Charles}
{Charles University, Faculty of Mathematics and Physics,
 Institute of Particle and Nuclear Physics, Prague, Czech Republic}
\newcommand{\Cincinnati}{Department of Physics, University of Cincinnati, 
Cincinnati, Ohio 45221, USA}
\newcommand{\CSU}{Department of Physics, Colorado 
State University, Fort Collins, CO 80523-1875, USA}
\newcommand{\CTU}{Czech Technical University in Prague,
Brehova 7, 115 19 Prague 1, Czech Republic}
\newcommand{\Dallas}{Physics Department, University of Texas at Dallas,
800 W. Campbell Rd. Richardson, Texas 75083-0688, USA}
\newcommand{\DallasU}{University of Dallas, 1845 E 
Northgate Drive, Irving, Texas 75062 USA}
\newcommand{\Delhi}{Department of Physics and Astrophysics, University of 
Delhi, Delhi 110007, India}
\newcommand{\JINR}{Joint Institute for Nuclear Research,  
Dubna, Moscow region 141980, Russia}
\newcommand{\Erciyes}{
Department of Physics, Erciyes University, Kayseri 38030, Turkey}
\newcommand{\FNAL}{Fermi National Accelerator Laboratory, Batavia, 
Illinois 60510, USA}
\newcommand{\FSU}{Florida State University, Tallahassee, Florida 32306, USA}
\newcommand{\UFG}{Instituto de F\'{i}sica, Universidade Federal de 
Goi\'{a}s, Goi\^{a}nia, Goi\'{a}s, 74690-900, Brazil}
\newcommand{\Guwahati}{Department of Physics, IIT Guwahati, Guwahati, 781 
039, India}
\newcommand{\Harvard}{Department of Physics, Harvard University, 
Cambridge, Massachusetts 02138, USA}
\newcommand{\Houston}{Department of Physics, 
University of Houston, Houston, Texas 77204, USA}
\newcommand{\IHyderabad}{Department of Physics, IIT Hyderabad, Hyderabad, 
502 205, India}
\newcommand{\Hyderabad}{School of Physics, University of Hyderabad, 
Hyderabad, 500 046, India}
\newcommand{\IIT}{Illinois Institute of Technology,
Chicago IL 60616, USA}
\newcommand{\Indiana}{Indiana University, Bloomington, Indiana 47405, 
USA}
\newcommand{\INR}{Institute for Nuclear Research of Russia, Academy of 
Sciences 7a, 60th October Anniversary prospect, Moscow 117312, Russia}
\newcommand{\Iowa}{Department of Physics and Astronomy, Iowa State 
University, Ames, Iowa 50011, USA}
\newcommand{\Irvine}{Department of Physics and Astronomy, 
University of California at Irvine, Irvine, California 92697, USA}
\newcommand{\Jammu}{Department of Physics and Electronics, University of 
Jammu, Jammu Tawi, 180 006, Jammu and Kashmir, India}
\newcommand{\Lebedev}{Nuclear Physics and Astrophysics Division, Lebedev 
Physical 
Institute, Leninsky Prospect 53, 119991 Moscow, Russia}
\newcommand{\Magdalena}{Universidad del Magdalena, Carrera 32 No 22-08 Santa Marta, Colombia}
\newcommand{\MSU}{Department of Physics and Astronomy, Michigan State 
University, East Lansing, Michigan 48824, USA}
\newcommand{\Crookston}{Math, Science and Technology Department, University 
of Minnesota Crookston, Crookston, Minnesota 56716, USA}
\newcommand{\Duluth}{Department of Physics and Astronomy, 
University of Minnesota Duluth, Duluth, Minnesota 55812, USA}
\newcommand{\Minnesota}{School of Physics and Astronomy, University of 
Minnesota Twin Cities, Minneapolis, Minnesota 55455, USA}
\newcommand{\Mississippi}{University of Mississippi, University, Mississippi 38677, USA}
\newcommand{\NISER}{National Institute of Science Education and Research,
Khurda, 752050, Odisha, India}
\newcommand{\Oxford}{Subdepartment of Particle Physics, 
University of Oxford, Oxford OX1 3RH, United Kingdom}
\newcommand{\Panjab}{Department of Physics, Panjab University, 
Chandigarh, 160 014, India}
\newcommand{\Pitt}{Department of Physics, 
University of Pittsburgh, Pittsburgh, Pennsylvania 15260, USA}
\newcommand{\QMU}{Particle Physics Research Centre, 
Department of Physics and Astronomy,
Queen Mary University of London,
London E1 4NS, United Kingdom}
\newcommand{\RAL}{Rutherford Appleton Laboratory, Science 
and 
Technology Facilities Council, Didcot, OX11 0QX, United Kingdom}
\newcommand{\SAlabama}{Department of Physics, University of 
South Alabama, Mobile, Alabama 36688, USA} 
\newcommand{\Carolina}{Department of Physics and Astronomy, University of 
South Carolina, Columbia, South Carolina 29208, USA}
\newcommand{\SDakota}{South Dakota School of Mines and Technology, Rapid 
City, South Dakota 57701, USA}
\newcommand{\SMU}{Department of Physics, Southern Methodist University, 
Dallas, Texas 75275, USA}
\newcommand{\Stanford}{Department of Physics, Stanford University, 
Stanford, California 94305, USA}
\newcommand{\Sussex}{Department of Physics and Astronomy, University of 
Sussex, Falmer, Brighton BN1 9QH, United Kingdom}
\newcommand{\Syracuse}{Department of Physics, Syracuse University,
Syracuse NY 13210, USA}
\newcommand{\Tennessee}{Department of Physics and Astronomy, 
University of Tennessee, Knoxville, Tennessee 37996, USA}
\newcommand{\Texas}{Department of Physics, University of Texas at Austin, 
Austin, Texas 78712, USA}
\newcommand{\Tufts}{Department of Physics and Astronomy, Tufts University, Medford, 
Massachusetts 02155, USA}
\newcommand{\UCL}{Physics and Astronomy Department, University College 
London, 
Gower Street, London WC1E 6BT, United Kingdom}
\newcommand{\Virginia}{Department of Physics, University of Virginia, 
Charlottesville, Virginia 22904, USA}
\newcommand{\WSU}{Department of Mathematics, Statistics, and Physics,
 Wichita State University, 
Wichita, Kansas 67260, USA}
\newcommand{\WandM}{Department of Physics, William \& Mary, 
Williamsburg, Virginia 23187, USA}
\newcommand{\Wisconsin}{Department of Physics, University of 
Wisconsin-Madison, Madison, Wisconsin 53706, USA}
\newcommand{\deceased}{Deceased.}
\affiliation{\ANL}
\affiliation{\Atlantico}
\affiliation{\Bandirma}
\affiliation{\BHU}
\affiliation{\Caltech}
\affiliation{\Charles}
\affiliation{\Cincinnati}
\affiliation{\Cochin}
\affiliation{\CSU}
\affiliation{\CTU}
\affiliation{\Delhi}
\affiliation{\Erciyes}
\affiliation{\FNAL}
\affiliation{\FSU}
\affiliation{\UFG}
\affiliation{\Guwahati}
\affiliation{\Houston}
\affiliation{\Hyderabad}
\affiliation{\IHyderabad}
\affiliation{\IIT}
\affiliation{\Indiana}
\affiliation{\ICS}
\affiliation{\INR}
\affiliation{\IOP}
\affiliation{\Iowa}
\affiliation{\Irvine}
\affiliation{\JINR}
\affiliation{\Magdalena}
\affiliation{\MSU}
\affiliation{\Duluth}
\affiliation{\Minnesota}
\affiliation{\Mississippi}
\affiliation{\NISER}
\affiliation{\Panjab}
\affiliation{\Pitt}
\affiliation{\QMU}
\affiliation{\SAlabama}
\affiliation{\Carolina}
\affiliation{\SMU}
\affiliation{\Sussex}
\affiliation{\Syracuse}
\affiliation{\Texas}
\affiliation{\Tufts}
\affiliation{\UCL}
\affiliation{\Virginia}
\affiliation{\WSU}
\affiliation{\WandM}
\affiliation{\Wisconsin}

\author{M.~A.~Acero}
\affiliation{\Atlantico}

\author{B.~Acharya}
\affiliation{\Mississippi}

\author{P.~Adamson}
\affiliation{\FNAL}



\author{L.~Aliaga}
\affiliation{\FNAL}






\author{N.~Anfimov}
\affiliation{\JINR}


\author{A.~Antoshkin}
\affiliation{\JINR}


\author{E.~Arrieta-Diaz}
\affiliation{\Magdalena}

\author{L.~Asquith}
\affiliation{\Sussex}


\author{A.~Aurisano}
\affiliation{\Cincinnati}


\author{A.~Back}
\affiliation{\Indiana}



% ...

\collaboration{The NOvA Collaboration}
\noaffiliation

\date{\today}

